\documentclass[secnumarabic,amssymb, nobibnotes, aps, prd]{revtex4-1}
\usepackage{amsmath} \usepackage{amssymb}
\usepackage{graphicx} 
\usepackage{epstopdf}
\usepackage{amsmath}
\usepackage{amsmath,amssymb,amsthm,amsfonts,mathrsfs,bm,verbatim}
\usepackage{graphicx,subfigure}

\newcommand{\bea}{\begin{eqnarray}}
\newcommand{\eea}{\end{eqnarray}}
\def\nn{\nonumber}

  \def\D{\Delta}

\def\l{\lambda}

\def\o{\omega}  \def\O{\Omega}

\def\th{\theta}
\begin{document}

\title{Higher-dimensional non-extremal Reissner-Nordstrom black holes, scalar perturbation and superradiance: an analytical study}

\author{Jia-Hui Huang}
\email{huangjh@m.scnu.edu.cn}
\affiliation{Guangdong Provincial Key Laboratory of Quantum Engineering and Quantum Materials,
School of Physics and Telecommunication Engineering,
South China Normal University, Guangzhou 510006, China}
\affiliation{Institute of quantum matter,South China Normal University, Guangzhou 510006, China}
\affiliation{Guangdong-Hong Kong Joint Laboratory of Quantum Matter, Southern Nuclear Science Computing Center, South China Normal University, Guangzhou 510006, China}

\author{Run-Dong Zhao,Yi-Feng Zou}
\affiliation{Guangdong Provincial Key Laboratory of Nuclear Science, Institute of quantum matter,South China Normal University, Guangzhou 510006, China}
\affiliation{Guangdong-Hong Kong Joint Laboratory of Quantum Matter, Southern Nuclear Science Computing Center, South China Normal University, Guangzhou 510006, China}

\begin{abstract}
The superradiant stability of higher dimensional non-extremal Reissner-Nordstrom black holes under charged massive scalar perturbation is analytically studied. We extend an analytical method developed by one of the authors in the extremal Reissner-Nordstrom black hole cases to non-extremal cases. Using the new analytical method, we revisit four-dimensional Reissner-Nordstrom black hole case and obtain that four-dimensional Reissner-Nordstrom black hole is superradiantly stable, which is consistent with results in previous works. We then analytically prove that the five-dimensional Reissner-Nordstrom black holes are also superradiantly stable under charged massive scalar perturbation. Our result implies that all higher dimensional non-extremal Reissner-Nordstrom black holes may be superradiantly stable under charged massive scalar perturbation.
\end{abstract}

\maketitle

\section{Introduction}
Black holes are mysterious objects in spacetime predicted by general relativity. It is well-known that nothing can escape from a black hole when falling into its horizon.
However, it was suggested that energy might be extracted from a black hole by observers outside the black hole in early 1970's \cite{Penrose:1971uk,Bardeen:1972fi,Press:1972zz,Bekenstein:1973mi}.
When a charged bosonic wave is scattering off a charged rotating black hole, the wave is amplified by the black hole if the wave frequency $\omega$ obeys
  \begin{equation}\label{superRe}
   \omega < n\Omega_H  + e\Phi_H,
  \end{equation}
where $e$ and $n$ are the charge and azimuthal number of the bosonic wave mode, $\Omega_H$ is the angular velocity of the black hole horizon and $\Phi_H$ is the electromagnetic potential of the black hole horizon. This kind of wave amplification process is called superradiant scattering, which in fact has broad applications in various areas of physics(for a recent review, see\cite{Brito:2015oca}).

When there is a mirror-like mechanism that makes the amplified wave be scattered back and forth between the mirror and the black hole, the background black hole geometry will become superradiantly unstable. This is dubbed the black hole bomb mechanism \cite{Press:1972zz,Cardoso:2004nk,Herdeiro:2013pia,Degollado:2013bha}. For the massive bosonic perturbation, its  mass term behaves as a natural mirror. The superradiant (in)stability of asymptotically flat rotating black holes under massive scalar and vector perturbation has been studied extensively in the literature \cite{Strafuss:2004qc,Konoplya:2006br,Cardoso:2011xi,Dolan:2012yt,Hod:2012zza,Hod:2014pza,Aliev:2014aba,Hod:2016iri,Degollado:2018ypf,Lin:2021ssw,Xu:2020fgq,Huang:2019xbu,Ponglertsakul:2020ufm,East:2017ovw,East:2017mrj}.

Four-dimensional asymptotically flat charged Reissner-Nordstrom (RN) black holes were proved superradiantly stable against charged massive scalar perturbation \cite{Hod:2013eea,Huang:2015jza,Hod:2015hza,DiMenza:2014vpa,Chowdhury:2019ptb}. The logic in the proofs is as follows: the effect of the curved black hole geometry on the motion of the scalar perturbation can be described by an effective potential outside the black hole horizon and when superradiant modes exist for a charged massive scalar perturbation in a RN black hole background, there is no trapping potential well for the effective potential outside the black hole horizon, which could reflect the superradiant modes back and forth \cite{Li:2014gfg,Sanchis-Gual:2015lje,Fierro:2017fky,Li:2014fna,Li:2015mqa}. Higher dimensional RN black holes has also been studied numerically in literature \cite{Konoplya:2011qq,Konoplya:2007jv,Konoplya:2008au,Konoplya:2013sba,Konoplya:2008rq,Kodama:2003kk,Kodama:2007sf,Ishibashi:2011ws,Ishihara:2008re,Ishibashi:2003ap}.
Asymptotically flat RN black holes in $D=5,6,..,11$ are shown to be stable by studying the time-domain evolution of
the scalar perturbation with a numerical integration method \cite{Konoplya:2008au}.

Recently, an analytical method has been developed by one of the authors in the study of superradiant stability of higher dimensional extremal RN black holes \cite{Huang:2021dpa}. Using this analytical method, five and six-dimensional extremal RN black holes are found to be superradiantly stable under charged massive scalar perturbation. In this paper,
we  extend this  analytical method to non-extremal RN black hole cases.  First, we revisit the case of four-dimensional non-extremal RN black hole under scalar perturbation. Then we extend the analytical method to five-dimensional non-extremal RN black hole case and provide a systematic and analytical study of the superradiant stability of the black hole.

The organization of this paper is as follows. In Section II, we revisit the superradiant stability of four-dimensional RN black hole under a scalar perturbation with the new analytical method. The new method is systematic and appears simpler than previous analytical methods. In Section III, we first give a description of the motion of scalar perturbation on five-dimensional RN black hole background. Then we extend the new analytical method and provide a systematic and analytical study  for five-dimensional non-extremal RN black hole case. We obtain that the five-dimensional non-extremal RN black holes are superradiantly stable under charged scalar perturbation. The last Section is devoted to  a summary.

\section{Revisit of D=4 RN black holes}
In this section, we revisit the superradiant stability of a four-dimensional non-extremal RN black hole under charged massive scalar perturbation\cite{Huang:2015jza,Hod:2015hza}.
Here, we extend an analytical method developed in Ref.\cite{Huang:2021dpa} to the non-extremal RN black hole case. With the new method, we obtain that four-dimensional Reissner-Nordstrom black hole is superradiantly stable under charged massive scalar perturbation, which is the same as previous results \cite{Huang:2015jza,Hod:2015hza}.

The metric of four-dimensional non-extremal RN black hole is
\bea
ds^2=-f(r)dt^2+\frac{dr^2}{f(r)}+r^2d\Omega_{2}^2,
\eea
where
\bea
f(r)=1-\frac{2m}{r}+\frac{q^2}{r^2}.
\eea
$m$ and $q$ are respectively the mass and electric charge of the RN black hole. The inner and outer horizons of the black hole are
$r_\pm=m\pm\sqrt{m^2-q^2}$. It is obvious that
\bea
r_{+}r_{-}=q^{2};\quad r_{+}+r_{-}=2m.
\label{key1}
\eea
The background electromagnetic potential is
\bea
A_\mu=(-q/r,0,0,0)
\eea
The dynamics of a charged massive scalar field perturbation $\Psi(x)$ with mass $\mu$ and charge $e$ is governed by the Klein-Gordon equation,
\bea
(D_\nu D^\nu-\mu^2)\Psi=0,
\eea
where $D_\nu=\nabla_\nu-ie A_\nu$ is the covariant derivative. The solution of the above equation with definite angular frequency can be written as
\bea
\Psi=\sum_{lm} R_{lm}(r) Y_{lm}(\th,\phi) e^{-i\o t}.
\eea
$Y_{lm}$ is the spherical harmonic function, $l$ is the spherical harmonic index, $m$ is the azimuthal harmonic index with $-l\leqslant m\leqslant l$. The radial equation of motion satisfied by $R_{lm}$ is \cite{Huang:2015jza,Hod:2015hza},
\bea
\D \frac{d}{dr}(\D\frac{d R_{lm}}{dr})+ U R_{lm}=0,
\eea
where $\D=r^2-2m r+q^2$ and
\bea
U=(\o r^2-e q r)^2-\D[\mu^2r^2+l(l+1)].
\eea
By studying the asymptotic solutions of the radial equation with appropriately chosen boundary conditions, i.e. purely ingoing waves at the horizon and exponentially decaying states (bound states) at spatial infinity, we can obtain the superradiance condition and bound state condition \cite{Huang:2015jza,Hod:2015hza},
\bea\label{4Dconds}
0<\o<e q/r_+,~~ \o<\mu.
\eea

Defining a new radial function $\psi=\D^{1/2} R_{lm}$, the above radial equation of motion can be rewritten as a Schrodinger-like equation
\bea
\frac{d^2}{dr^2}\psi+(\o^2-V)\psi=0,
\eea
where the effective potential $V$ is
\bea
V=\o^2-\frac{U+m^2-q^2}{\D^2}
\eea
In order to see if there exists a trapping potential well outside the horizon, we could analyze the derivative of the effective potential $V$. The asymptotic behaviors of the potential $V$ near outer horizon and at spatial infinity are
\bea\label{4Dasym}\nn
V(r\rightarrow r_+)&\rightarrow& -\infty;\\
V(r\rightarrow +\infty)&\rightarrow& \mu^2+\frac{2m\mu^2+2eq\o-4m\o^2}{r}+O(\frac{1}{r^2}).
\eea
When $\o$ satisfies the superradiant condition and bound state condition \eqref{4Dconds}, we can obtain
\bea
2m\mu^2+2eq\o-4m\o^2=2m(\mu^2-\o^2)+2\o(e q-m\o)>0.
\eea
In the above, we use $\o<eq/r_+<eq/m$.
From the asymptotic behaviors of the effective potential \eqref{4Dasym}, we can know that there is at least one maximum for $V$ outside the horizon $r_+$.

In order to show that there is no trapping potential well for the effective potential $V$ outside the  horizon $r_+$, we will consider the extrema of the effective potential $V(r)$ by analyzing the real roots of the equation $V'(r)=0$ for $r>r_+$.

The derivative of effective potential $V$ is
\bea
V'(r)=-\frac{1}{\D^3}f(r),
\eea
where the numerator $f(r)$ is a polynomial of $r$. The explicit expression of $f(r)$ is
\bea
f(r)&=&a_{0}+a_{1}r+a_{2}r^{2}+a_{3}r^{3}+a_{4}r^{4},\label{key2}\\
a_{0}&=&4 m^3-4 m q^2-2  m q^2 \lambda_l ,\nn\\
a_{1}&=&-2 q^4 \mu ^2 +4 q^2-4 m^2+2 q^4 e^2 +4 m^2 \lambda_l+2 q^2 \lambda_l  ,\nn\\
a_{2}&=&-6 q^3 e  \omega +6 m q^2 \mu ^2 -6m \lambda_l ,\nn\\
a_{3}&=&4 q^2 \omega ^2+4 m q  e \omega -4  m^2 \mu ^2-2 q^2 \mu ^2-2 q^2 e^2+2 \lambda_l,\nn\\
a_{4}&=& -4 m \omega ^2+2 q e \omega +2 m \mu ^2,
\eea
where $\lambda_l=l(l+1)$ is the eigenvalue of the angular equation of motion. Because we are interested in the real roots of $V'(r)=0$, we can instead consider the real roots of its numerator, i.e., the real roots of $f(r)=0$.

By defining a new variable $x=r-r_{+}$, we can rewrite the numerator of the derivative of the effective potential \eqref{key2}  as
\bea\label{g4D}
g(x)=f(r)=b_{0}+b_{1}x+b_{2}x^{2}+b_{3}x^{3}+b_{4}x^{4},
\eea
where
\bea
b_{0}&=&\frac{1}{2}\left(r_--r_+\right) [(r_+-r_-)^2+4r_+^2\left( e q-\o r_+\right)^2],\\
b_{1}&=&-4 r_{+}^{3}(2 r_{+}-r_{-})\omega^{2}+2 e q r_{+}^{2}(7 r_{+}-3r_{-}) \omega + (r_{+}-r_{-})^{2}r_{+}^{2}\mu^{2}\nn\\
&&+(r_{+}-r_{-})^{2}\lambda_l +2e^{2}q^2 r_+(r_{-}-3 r_{+})-(r_{+}-r_{-})^{2},\\
b_{2}&=&-12r_{+}^{3}\omega^{2}+18 e q r_{+}^{2}\omega
+3r_{+}^{2}(r_{+}-r_{-})\mu^{2}+3(r_{+}-r_{-})\lambda_l-6e^{2}q^2r_{+},\\
b_{3}&=&-(8 r_+^2+4 r_+ r_-)\omega^{2}+2 e q(5r_+ +r_-)\omega+(3 r_+^2 - r_-^2)\mu^{2}+2\lambda_l-2 e^{2}q^2,\\
b_{4}&=&2 e q \omega + (r_++r_-) (\mu^2 - 2 \omega^2).
\eea
Next, let's analyze the signs of the above coefficients. It is easy to see that $b_0<0$. According to the superradiance condition and bound state condition \eqref{4Dconds}, we can also easily obtain
\bea\nn
b_4&=&2 e q \omega-(r_{+}+ r_{-})\o^2 + (r_{+}+ r_{-})(\mu^2-\o^2)\\
&>&2 r_+\o^2-(r_{+}+ r_{-})\o^2 + (r_{+}+ r_{-})(\mu^2-\o^2)>0.
\eea
It is not easy to directly analyze the signs of $b_3, b_2, b_1$. We define three new quantities
\bea
b'_3=\frac{b_3}{8r_+^2+4r_+r_-},~
b'_2=\frac{b_2}{12r_+^3},~
b'_1=\frac{b_1}{4r_+^3(2r_+-r_-)}.
\eea
An important fact is that the sign of $b'_i(i=1,2,3)$ is the same as that of $b_i$. We will use this fact later. Let's first compute the difference between $b'_2$ and
$b'_1$,
\bea
b'_2-b'_1=\frac{1}{4r_+^3(2r_+-r_-)}[r_+(r_+-r_-)\l_l+(r_+-r_-)^2+\mu^2r_+^3(r_+-r_-)+2eqr_+^2(e q-\o r_+)].
\eea
Given the superradiance condition, $e q>\o r_+$, and $r_+>r_-$, we have
\bea\label{b2b1}
b'_2>b'_1.
\eea
Then let's compute the difference between $b'_3$ and $b'_2$,
\bea
b'_3-b'_2&=&\frac{1}{4r_+^3(2r_++r_-)}[r_-(r_++r_-)\l_l+(\mu^2-\o^2) r_+^2 (r_+ + r_-)\nn\\
&+&r_+ (2 e^2 q^2 (r_+ + r_-) + \o^2 r_+^2 (r_+ + r_-) -2 e q r_+ (2 r_- + r_+) \o)].
\eea
The $\l_l$ term in the square bracket is obviously positive and given the bound state condition, the $(\mu^2-\o^2)$ term is also obviously positive. The left in the square bracket is a quadratic function of $\o$ (ignoring the overall $r_+$),
\bea
f_1(\o)=  r_+^2 (r_+ + r_-)\o^2 -2 e q r_+ (2 r_- + r_+) \o+2 e^2 q^2 (r_+ + r_-).
\eea
The intercept of $f_1$ is obviously positive. The symmetric axis of $f_1$ is located at
\bea
\o_s=\frac{2 e q r_+ (2 r_- + r_+)}{ 2r_+^2 (r_+ + r_-)}=\frac{eq}{r_+}\frac{2 r_- + r_+}{r_+ + r_-}.
\eea
It is obvious that $\o_s>eq/r_+$. When the angular frequency satisfies the superradiance condition $0<\o<eq/r_+$, we can know that
\bea
f_1(\o)>f_1(eq/r_+)=(r_+-r_-)e^2q^2>0.
\eea
So given the superradiance condition and bound state condition, we can obtain
\bea\label{b3b2}
b'_3>b'_2.
\eea

According to equations \eqref{b2b1} \eqref{b3b2}, we have
\bea
b'_3>b'_2>b'_1.
\eea
The possible signs of ordered quantities $(b'_3, b'_2, b'_1)$ are
\bea
(+,+,+),~(+,+,-),~(+,-,-),~(-,-,-).
\eea
Using the fact that the sign of $b'_i(i=1,2,3)$ is the same as that of $b_i$,
so the possible signs of the ordered coefficients $(b_3, b_2, b_1)$ are
\bea
(+,+,+),~(+,+,-),~(+,-,-),~(-,-,-).
\eea
The possible signs of the ordered coefficients $(b_4,b_3, b_2, b_1,b_0)$ are
\bea\label{4dsign}
(+,+,+,+,-),~(+,+,+,-,-),~(+,+,-,-,-),~(+,-,-,-,-).
\eea
The sign change of the ordered coefficients $(b_4,b_3, b_2, b_1,b_0)$ is always 1. According to \textit{Descartes' rule of signs}, we know that there is at most one positive real root for equation $g(x)=0$, defined in \eqref{g4D}. Namely, there is at most one real root for equation $f(r)=0$ when $r>r_+$. From the asymptotic analysis of the effective potential in \eqref{4Dasym}, we know that there is one maximum when $r>r_+$. So there is no potential well (minimum)for the effective potential $V(r)$ when $r>r_+$. We conclude that the four-dimensional RN black hole is superradiantly stable under charged massive scalar perturbation.

\section{D=5 RN black holes}
In this section, we analytically study the superradiant stability of D=5 RN black hole under charged massive scalar perturbation. It is proved that given the superradiance condition and bound state condition, there is no trapping potential well outside the horizon for the effective potential experienced by the scalar perturbation. The black hole and scalar perturbation system is superradiantly stable. For simplicity, we use many symbols which are the same as the D=4 case.

The metric of  D=5 RN black hole is
\bea
ds_5^2=-f(r)dt^2+\frac{dr^2}{f(r)}+r^2d\O_3^2,
\eea
where
\bea
f(r)=1-\frac{2m}{r^2}+\frac{q^2}{r^{4}}.
\eea
The parameters $m$ and $q$ are related with the ADM mass $M$ and electric charge $Q$ of the black hole \cite{Myers:1986un,Destounis:2019hca},
\bea
m=\frac{4}{3\pi}M,~~q=\frac{2}{\sqrt{3}\pi}Q.
\eea
The inner and outer horizons are
\bea
r_\pm=(m\pm\sqrt{m^2-q^2})^{1/2}.
\eea
It is obvious that the inner and outer horizons satisfy the following relations
\bea\label{5DmqRpm}
r_+r_-=q,~~ r_+^2+r_-^2=2m.
\eea
The background electromagnetic potential is
\bea
A_\mu=(-\frac{\sqrt{3}q}{2r^2},0,0,0,0).
\eea
The dynamics of a charged massive scalar field perturbation $\Psi(x)$ with mass $\mu$ and charge $e$ is governed by the Klein-Gordon equation,
\bea
(D_\nu D^\nu-\mu^2)\Psi=0,
\eea
where $D_\nu=\nabla_\nu-ie A_\nu$ is the covariant derivative. Similar with the D=4 case, the above equation of motion can be separated into radial part and angular part. The eigenfunctions of the angular equation of motion are scalar harmonics on $S^3$ with eigenvalue $\l_l=l(l+2),(l\geqslant 0)$ \cite{Chodos:1983zi,Higuchi:1986wu,Rubin1984,Achour:2015zpa,Lindblom:2017maa}. The radial equation of motion is
\bea\label{5Dradial}
\Delta\frac{d}{dr}(\Delta\frac{d R}{dr})+U R=0,
\eea
where $R$ is the radial function,  $\Delta=r^{3}f(r)$ and
\bea
U=(\o-\frac{\sqrt{3}}{2}\frac{eq}{r^2})^2 r^{6}-l(l+2) r\Delta-\mu^2 r^{3}\Delta.
\eea

In order to study the superradiant stability of RN black hole under the charged massive scalar perturbation, appropriate boundary conditions should be considered for asymptotic solutions of the radial equation near the horizon and at spatial
infinity. Define the tortoise coordinate $y$ by $dy=f^{-1}dr$ and a new radial function $\tilde{R}=r^{\frac{3}{2}}R$, then
the radial equation \eqref{5Dradial} can be rewritten as
\bea
\frac{d^2\tilde{R}}{dy^2}+\tilde{U} \tilde{R}=0,
\eea
where
\bea
\tilde{U}=\frac{U}{r^{6}}-\frac{3f(r)[f(r)+ 2 r f'(r)]}{4r^2}.
\eea
The asymptotic behaviors of $\tilde{U}$ at the spatial infinity and outer horizon are
\bea
\lim_{r\rightarrow +\infty}\tilde{U}= \o^2-\mu^2,~~
\lim_{r\rightarrow r_+} \tilde{U}= (\o-\frac{\sqrt{3}}{2}\frac{e q}{r_+^2})^2=(\o-e\phi_H)^2,
\eea
where $\phi_H$ is the electric potential of the outer horizon of the RN black hole. We need ingoing wave condition near the outer horizon and bound state condition at spatial infinity. Then the asymptotic solutions of the radial wave equation are chosen as the following
\bea
y\to +\infty (r\to +\infty ),~~~ \tilde{R}\sim {e}^{-\sqrt{{{\mu }^{2}}-{{\omega }^{2}}}y},\\
y\to -\infty ( r\to r_+),~~~\tilde{R}\sim {{e}^{-i(\omega -e{{\phi }_{H}})y}}.
\eea
It is easy to see that bound state condition at spatial infinity requires the following inequality
\bea\label{bound-con}
\o<\mu.
\eea
The superradiance condition in this case is
\bea\label{sup-con}
0<\o<\o_c=e\phi_H=\frac{\sqrt{3}}{2}\frac{e q}{r_+^{2}}.
\eea

\subsection{Effective potential and asymptotic analysis}
\label{asmAna}
By defining a new radial function $\psi=\D^{1/2} R$, the radial equation of motion \eqref{5Dradial} can be rewritten as a Schrodinger-like equation
\bea
\frac{d^2}{dr^2}\psi+(\o^2-V)\psi=0,
\eea
where the effective potential $V$ is
\bea
V=\o^2+\frac{B}{A},
\eea
and
\bea
A&=&4r^2(r^4-2m r^2+q^2)^2,\\\nn
B&=&(4 \mu^2 - 4 \o^2)r^{10}+(3 - 8 m \mu^2 + 4 \sqrt{3} e q \o + 4 \lambda_l)r^8 +
 (-12 m - 3 e^2 q^2 + 4 \mu^2 q^2 - 8 m \lambda_l)r^6 \\
 &+& (-4 m^2 + 22 q^2 + 4 q^2 \lambda_l) r^4- 12 m q^2 r^2+3 q^4.
\eea
The asymptotic behaviors of the effective potential $V$ near outer horizon and at spatial infinity are
\bea\label{5Dasym}\nn
V(r\rightarrow r_+)&\rightarrow& -\infty;\\
V(r\rightarrow +\infty)&\rightarrow& \mu^2+\frac{3/4+2m\mu^2+\sqrt{3}eq\o-4m\o^2+\l_l}{r^2}+O(\frac{1}{r^4}).
\eea
When $\o$ satisfies the superradiant condition \eqref{sup-con} and bound state condition \eqref{bound-con}, one can get
\bea
2m\mu^2+\sqrt{3}eq\o-4m\o^2=2m(\mu^2-\o^2)+2\o(\frac{\sqrt{3}}{2}e q-m\o)>0.
\eea
In the above, we use $\o<\frac{\sqrt{3}}{2}\frac{eq}{r_+^2}<\frac{\sqrt{3}}{2}\frac{eq}{m}$.

From the asymptotic behaviors of the effective potential \eqref{5Dasym}, we know that there is at least one maximum for the effective potential $V$ outside the outer horizon $r_+$.

\subsection{Analysis of derivative of the effective potential}

In this subsection we study if there exists a trapping potential well outside the horizon $r_+$ for the effective potential $V$. We do the study by analyzing the derivative of the effective potential, $V'(r)$. If there is only one real root for $V'=0$ when $r>r_+$, this root corresponds to the maximum discussed in the above subsection and there is no potential well for $V$ when  $r>r_+$.

The derivative of the effective potential $V$ is
\bea
V'(r)=-\frac{C_5}{2r^3\D^3},
\eea
where the numerator of derivative of effective potential $V$ is
\bea
C_5&=&r^{12} (-16 m \omega ^2+4 \sqrt{3} e q \omega +8 m \mu ^2+4 \lambda_l +3)\nn\\&+& r^{10} (16 q^2 \omega ^2+8 \sqrt{3}  m q e \omega - 8\left(2 m^2+ q^2\right) \mu ^2-6 q^2 e^2 -18 m-8 m \lambda_l )\nn\\
&+& r^8 (-12 \sqrt{3}  q^3 e \omega+24m q^2 \mu ^2 -12 m^2 +57 q^2)\nn\\&+& r^6 (-8 q^4 \mu ^2+8 m^3-68 m q^2+6 q^4 e^{2}+8 m q^2 \lambda_l)\nn\\ &+& r^4 (52 m^2 q^2 -7 q^4 + 4 q^{4} \lambda_l)+r^2(-18 m q^4) +3 q^6\nn\\
&=& r^{12}A_6+r^{10}A_5+r^8A_4+r^6A_3+r^4A_2+r^2A_1+A_0.
\eea
Because we are interested in the real roots of $V'(r)=0$ when $r>r_+$, we can ignore the nonzero denominator of $V'(r)$ and consider the roots of $C_5(r)=0$ when $r>r_+$.
Making a change of variable $y=r^2-r_+^2$, $C_5$ can be rewritten as
\bea\label{c5y}
C_{5}=B_{6}y^{6}+B_{5}y^{5}+B_{4}y^{4}+B_{3}y^{3}+B_{2}y^{2}+B_{1}y+B_{0},
\eea
where
\bea
B_{0}&=&-8 r_+^{12}\left(r_+^{2}-r_-^2\right) \omega ^2\nn+ 8 \sqrt{3}r_+^{10}\left(r_+^{2}  - r_-^2 \right)e q\omega -6 r_+^{8} (r_+^{2}- r_-^2)e^{2}q^2\nn\\
&-&8 r_+^6 ( r_+^{2}-r_-^2)^3,\\
B_{1}&=&-16\left(3 r_+^{12}-2 r_+^{10} r_-^2\right) \omega ^2+ 4 \sqrt{3}\left(11 r_+^{10}  -7r_+^8 r_-^2 \right)e q\omega+4 \left( r_+^{12}-2 r_+^{10} r_-^2+  r_+^8 r_-^4\right)\mu ^2 \nn\\
&-&6(5r_+^{8}  -3r_+^6 r_-^2)e^{2}q^2 -36 r_+^{10}+92r_+^8 r_-^2-76r_+^6 r_-^4+20r_+^4 r_-^6\nn\\
&+& 4\left(r_+^{10}-2 r_+^8 r_-^2+ r_+^6 r_-^4\right)\lambda_l, \\
B_{2}&=&-40\left(3 r_+^{10}- r_+^8 r_-^2 \right) \omega ^2+ 4 \sqrt{3} \left(25 r_+^8 -8 r_+^6 r_-^2\right)e q \omega+ 4\left(5 r_+^{10}-7 r_+^8 r_-^2+2 r_+^6 r_-^4\right)\mu ^2\nn\\
 &-&6\left(10 r_+^6 -3r_+^4 r_-^2 \right)e^2 q^2-60 r_+^8+136 r_+^6 r_-^2 -92 r_+^4 r_-^4+16r_+^2 r_-^6\nn\\
 &+&4\left(5 r_+^8-7r_+^6 r_-^2 +2 r_+^4 r_-^4\right)\lambda_l,\\
B_{3}&=&-160 r_+^8 \omega ^2+8 \sqrt{3} \left(15 r_+^6  - r_+^4 r_-^2\right)e q \omega+8 \left(5 r_+^8-4 r_+^6 r_-^2\right)\mu ^2\nn\\
&-&6\left(10 r_-^4+r_+^2 r_-^2\right)e^2 q^2-41 r_+^6+83 r_+^4 r_-^2-43 r_+^2 r_-^4+r_-^6+4\left(10 r_+^6-9 r_+^4 r_-^2 + r_+^2 r_-^4 \right)\lambda_l,\\
B_{4}&=&-40\left(3 r_+^6+ r_+^4 r_-^2\right) \omega ^2+ 8 \sqrt{3} \left(10 r_+^4  +  r_+^2 r_-^2\right)e q \omega+8\left(5 r_+^6-r_+^4 r_-^2-r_+^2 r_-^4 \right)\mu ^2\nn\\
 &-&3 (r_+^4-2 r_+^2 r_-^2+ r_-^4)-30 r_+^2  e^2 q^2+20( 2r_+^4 - r_+^2 r_-^2)\lambda_l,\\
B_{5}&=&-16 \left(3 r_+^4+2 r_{+}^{2} r_-^2\right)\omega^{2}+4 \sqrt{3} \left(7r_+^2 +r_-^2\right)e q\omega+4\left(5 r_+^4+2 r_+^2 r_-^2- r_-^4\right)\mu ^2\nn\\
&+&9( r_+^2- r_-^2)-6 e^2 q^2+4(5 r_{+}^{2}-r_{-}^{2})\lambda_l, \\
B_6&=&8 m(\mu^2-2\o^2)+4 \sqrt{3} e q\omega +4 \lambda_l+3.
\eea

In order to use the new method based on the Descartes' rule of signs to prove there is no potential well for the effective potential outside the horizon $r_+$, we analyze the signs or sign relations of the coefficients $B_i (i=0,..,6)$.

The coefficient $B_6$ can be rewritten as
\bea
B_6=8 m(\mu^2-\o^2)+8m\omega(\frac{\sqrt{3}}{2} \frac{e q}{m}-\o) +4 \lambda_l+3.
\eea
 Based on the superradiant condition \eqref{sup-con} and the black hole parameter relation \eqref{5DmqRpm}, we know $0<\o<\o_c=\frac{\sqrt{3}}{2} \frac{e q}{r_+^2}<\frac{\sqrt{3}}{2} \frac{e q}{m}$. And with the bound state condition \eqref{bound-con}, we have
\bea\label{B6}
B_6>0.
\eea
$B_0$ is obviously negative when it is rewritten as following
\bea\label{B0}
B_0=-2r_+^8(r_+^2-r_-^2)(2r_+^2\o-\sqrt{3}eq)^2-8 r_+^6 ( r_+^{2}-r_-^2)^3<0.
\eea

It is not easy to judge the signs of other coefficients. In the next, we study the sign relations between pairs of coefficients. First, we define several scaled new coefficients as
\bea
B'_5=\frac{B_5}{16 \left(3 r_+^4+2 r_{+}^{2} r_-^2\right)},
B'_4=\frac{B_4}{40\left(3 r_+^6+ r_+^4 r_-^2\right)},
B'_3=\frac{B_3}{160 r_+^8}, B'_2=\frac{B_2}{40(3r_+^{10}-r_+^8r_-^2)},
B'_1=\frac{B_1}{16(3r_+^{12}-2r_+^{10}r_-^2)}.
\eea
It is worth noting that the scaling factors are all positive. So $B_i$ and $B'_i$ are simultaneously positive or negative, i.e. they enjoy the same sign.
Consider the difference of $B'_5$ and $B'_4$,
\bea
&&B'_5-B'_4=\frac{B_5}{16 \left(3 r_+^4+2 r_{+}^{2} r_-^2\right)}-\frac{B_4}{40\left(3 r_+^6+ r_+^4 r_-^2\right)}
=\frac{1}{80 r_+^4 \left(9 r_+^4+9 r_+^2 r_-^2+2 r_-^4\right)}\nn\\&\times&[\left(20 r_+^8+36 r_+^6 r_-^2+20 r_+^4r_-^4+4 r_+^2 r_-^6\right)\omega^{2}-4\sqrt{3} \left(5 r_+^6 +14 r_+^4r_-^2 + r_+^2r_-^4 \right)e q\omega \nn\\
&+&\left(30r_+^4 +30 r_+^2 r_-^2\right)e^2 q^2+51 r_+^6-38 r_+^4 r_-^{2}-17 r_+^2 r_-^4+4 r_-^6+20 (r_+^6+r_+^2 r_-^4) \lambda_l\nn\\
&+&\left(20 r_+^8+36 r_+^6 r_-^2+20 r_+^4r_-^4+4 r_+^2 r_-^6\right)(\mu^{2}-\omega^{2})].
\eea
It is easy to see the denominator of the above equation is positive.
Given the bound state condition $\o^2<\mu^2$ and $\l_l\geqslant0$, it is also easy to see that the $(\mu^2-\o^2)$ term and  $\l_l$ term  in the numerator of the above equation are positive. We take the left terms in the numerator of the above equation as a quadratic function of $\o$, which is defined as
\bea
g_1(\omega)=a_1\omega^{2}+b_1\omega+c_1,
\eea
where
\bea
a_1&=&20 r_+^8+36 r_+^6 r_-^2+20 r_+^4r_-^4+4 r_+^2 r_-^6>0,\\
b_1&=&-4\sqrt{3} \left(5 r_+^6+14 r_+^4r_-^2 + r_+^2r_-^4 \right)e q<0,\\
c_1&=&\left(30r_+^4 +30 r_+^2 r_-^2\right)e^2q^2+51 r_+^6-38 r_+^4 r_-^{2}-17 r_+^2 r_-^4+4 r_-^6\nn\\
&=&(30r_+^4 +30 r_+^2 r_-^2) e^2q^2+38(r_+^6-r_+^4 r_-^2)+(r_+^2r_--r_-^3)(13r_+^2r_--4r_-^3)>0.
\eea
The symmetric axis of $g_1(\o)$ is located at
\bea
\o_{1s}&=&-\frac{b_1}{2a_1}=\frac{2\sqrt{3} \left(5 r_+^6+14 r_+^4r_-^2 + r_+^2r_-^4 \right)e q}{20 r_+^8+36 r_+^6 r_-^2+20 r_+^4r_-^4+4 r_+^2 r_-^6}\nn\\
&=&\frac{\sqrt{3}eq}{2r_+^2}\frac{5 r_+^6+14 r_+^4r_-^2 + r_+^2r_-^4 }{5 r_+^6+9 r_+^4 r_-^2+5 r_+^2r_-^4+ r_-^6}.
\eea
Given $r_+>r_-$, one can check that
\bea
\o_{1s}>\frac{\sqrt{3}eq}{2r_+^2}=\o_c.
\eea
When the angular frequency $\o$ satisfies the superradiance condition, $0<\o<\o_c$, we know that $g_1(\o)>g_1(\o_c)$. Now let's compute $g_1(\o_c)$,
\bea
g_1(\o_c)&=&\left(15 r_+^6 -27 r_+^4 r_-^2+9 r_+^2 r_-^4+3 r_-^6\right)e^2q^2/r_+^2+51 r_+^6-38 r_+^4 r_-^2-17 r_+^2 r_-^4+4 r_-^6\nn\\
&=&3(r_+^2-r_-^2)^2(5r_+^2+r_-^2)e^2q^2/r_+^2+38(r_+^6-r_+^4 r_-^2)+(r_+^2r_--r_-^3)(13r_+^2r_--4r_-^3)\nn\\
&>&0.
\eea
So we have $g_1(\o)>0$ when $0<\o<\o_c$. Finally, given the superradiant condition and the bound state condition, we obtain
\bea\label{B54}
B'_5>B'_4.
\eea

Then let's consider the difference of $B'_4$ and $B'_3$.
\bea
B'_4-B'_3=\frac{B_4}{40\left(3 r_+^6+ r_+^4 r_-^2\right)}-\frac{B_3}{160 r_+^8}
=\frac{1}{80r_+^8(3r_+^2+r_-^2)}\nn\\
\times [4 r_+^8 (5 r_+^2+3 r_-^2 ) \o^2-4 \sqrt{3}r_+^4 (5r_+^4 + 8 r_+^2 r_-^2 -r_-^4)eq \o\nn\\+e^2 q^2 (-3 r_-^4 r_+^2 + 21 r_-^2 r_+^4 + 30  r_+^6)
+\frac{1}{2} (r_+^2 - r_-^2)^2 (-r_-^4 + 38 r_-^2 r_+^2 + 111 r_+^4)\nn\\
+2 r_+^2 (10 r_+^6- 3 r_+^4 r_-^2+6r_+^2r_-^4-r_-^6)\l_l
+4 r_+^8 (5 r_+^2+3 r_-^2 )(\mu^2-\o^2)].
\eea
Given the bound state condition $\o^2<\mu^2$ and $r_+>r_-$, it is easy to see from the above equation that the denominator of the difference is positive, the $\l_l$ term in the numerator is positive and the $(\mu^2-\o^2)$ term is also positive.
The left terms in the numerator can also be treated as a quadratic function of $\o$, which is defined as
\bea
g_2(\o)=4 r_+^8 (5 r_+^2+3 r_-^2 ) \o^2-4 \sqrt{3}r_+^4 (5r_+^4 + 8 r_+^2 r_-^2 -r_-^4)eq \o\nn\\+e^2 q^2 (-3 r_-^4 r_+^2 + 21 r_-^2 r_+^4 + 30  r_+^6)
+\frac{1}{2} (r_+^2 - r_-^2)^2 (-r_-^4 + 38 r_-^2 r_+^2 + 111 r_+^4).
\eea
It is easy to see that the intercept of $g_2$ is positive. The symmetric axis of  $g_2$ is also positive, which is located at
\bea
\o_{2s}=\frac{\sqrt{3}eq }{2 r_+^2 }\frac{(5r_+^4 + 8 r_+^2 r_-^2 -r_-^4)}{(5 r_+^4+3 r_+^2r_-^2 )}.
\eea
It is obvious that $\o_{2s}>\o_c$. When $\o$ satisfies the superradiance condition, one can know that
\bea
g_2(\o)>g_2(\o_c)=\frac{1}{2}(r_+^2-r_-^2)[(30r_+^4-6r_+^2r_-^2)e^2q^2+(r_+^2-r_-^2)
(111r_+^4+38r_+^2r_-^2-r_-^4)]>0.
\eea
So we finally have
\bea\label{B43}
B'_4>B'_3.
\eea

Next, let's consider the difference of $B'_3$ and $B'_2$,
 \bea
B'_3-B'_2=\frac{B_3}{160 r_+^8}-\frac{B_2}{40(3r_+^{10}-r_+^8r_-^2)}
=\frac{1}{160 r_+^8 (-r_-^2 + 3 r_+^2)}\times
[8r_+^8(5r_+^2-3r_-^2)\o^2\nn\\-8 \sqrt{3} r_+^4 \left(-r_-^4+2 r_-^2 r_+^2+5 r_+^4\right)e q \o +6e^2q^2 r_+^2(10r_+^4+r_+^2r_-^2-r_-^4)\nn\\+(r_+^2-r_-^2)^2(117r_+^4-20r_+^2r_-^2-r_-^4)
+4r_+^2(2r_+^2-r_-^2)(5r_+^4-2r_+^2r_-^2+r_-^4)\l_l\nn\\
+8r_+^8(5r_+^2-3r_-^2)(\mu^2-\o^2)].
 \eea
It is easy to see that the denominator of the above equation is positive. The $\l_l$ term and the $(\mu^2-\o^2)$ term in the square bracket are also positive.
The left terms in the square bracket can be taken as a quadratic function of
$\o$, which is defined as
\bea
g_3(\o)=8r_+^8(5r_+^2-3r_-^2)\o^2-8 \sqrt{3} r_+^4 \left(r_-^4-2 r_-^2 r_+^2-5 r_+^4\right)e q \o\nn\\ +6e^2q^2 r_+^2(10r_+^4+r_+^2r_-^2-r_-^4)+(r_+^2-r_-^2)^2(117r_+^4-20r_+^2r_-^2-r_-^4).
\eea
It is easy to see that the intercept of $g_3$ is positive. The symmetric axis of  $g_3$ is also positive, which is located at
\bea
\o_{3s}=\frac{8 \sqrt{3} r_+^4 \left(-r_-^4+2 r_-^2 r_+^2+5 r_+^4\right)e q}{16r_+^8(5r_+^2-3r_-^2)}=\frac{\sqrt{3}eq}{2r_+^2}
\frac{5r_+^4+2r_+^2r_-^2-r_-^4}{5r_+^4-3r_+^2r_-^2}.
\eea
It is obvious that $\o_{3s}>\o_c$. When $\o$ satisfies the superradiance condition, one can know that
\bea
g_3(\o)>g_3(\o_c)=(r_+^2-r_-^2)(30r_+^4-6r_+^2r_-^2)e^2q^2+(r_+^2-r_-^2)^2
(117r_+^4-20r_+^2r_-^2-r_-^4)>0.
\eea
So we obtain
\bea\label{B32}
B'_3>B'_2.
\eea

Finally, let's consider the difference of $B'_2$ and $B'_1$,
\bea
B'_2-B'_1=\frac{B_2}{40(3r_+^{10}-r_+^8r_-^2)}-\frac{B_1}{16(3r_+^{12}-2r_+^{10}r_-^2)}
=\frac{3(5r_+^4-4r_+^2r_-^2+r_-^4)}{40r_+^6(9r_+^4-9r_+^2r_-^2+2r_-^4)}\nn\\ \times [2 \sqrt{3} e q r_+^4(\frac{\sqrt{3}eq}{2r_+^2}-\o)+6(r_+^2-r_-^2)^2+2r_+^4(r_+^2-r_-^2)\mu^2+
2r_+^2(r_+^2-r_-^2)\l_l].
\eea
Given the superradiant condition and $r_+>r_-$, it is easy to see the above difference is positive. So we have
\bea\label{B21}
B'_2>B'_1.
\eea

According to the inequalities \eqref{B54}\eqref{B43}\eqref{B32}\eqref{B21}, we know that
\bea
B'_5>B'_4>B'_3>B'_2>B'_1.
\eea
The possible signs of the ordered coefficients $ (B'_5,B'_4,B'_3,B'_2,B'_1)$ are all plus, all minus or some plus on the left with minus on the right, i.e.
\bea\label{signs}
(+,+,+,+,+),~(-,-,-,-,-),~(+,..,+,-,..,-).
\eea
Because the scaling factor between $B'_i$ and $B_i$ is positive, the possible signs of the ordered coefficients$ (B_5,B_4,B_3,B_2,B_1)$ are the same as that listed above. With the results on $B_6,B_0$ in equations \eqref{B6}\eqref{B0}, we conclude that the possible signs of the ordered coefficients$ (B_6,B_5,B_4,B_3,B_2,B_1,B_0)$ are of the following form
\bea\label{5dsign}
(+,+,+,+,+,+,-),(+,+,+,+,+,-,-),(+,+,+,+,-,-,-),\nn\\(+,+,+,-,-,-,-),(+,+,-,-,-,-,-),
(+,-,-,-,-,-,-),
\eea
i.e. the plus signs are always on the left of the minus signs.

Consider the numerator $C_5$ of the derivative of the effective potential in equation \eqref{c5y}.
It is a polynomial of $y$ with real coefficients $ (B_6,B_5,B_4,B_3,B_2,B_1,B_0)$.
According to the theorem of\textit{ Descartes' rule of signs} about real roots of polynomial equations and the result in \eqref{5dsign}, we obtain that the sign change of the coefficients $ (B_6,B_5,B_4,B_3,B_2,B_1,B_0)$ is always 1 and there is at most one positive real root for equation $C_5(y)=0$. This means that there is at most one extreme for the effective potential felt by the scalar perturbation outside the horizon $r_+$. Based on the asymptotic analysis of the effective potential in section \eqref{asmAna}, we already know there is one maximum for the effective potential outside the horizon $r_+$. So there is no potential well (minimum) for the effective potential outside the horizon $r_+$.

\section{Summary}
In this work, we analytically study the superradiant stability of the higher dimensional non-extremal RN black holes under charged massive scalar perturbation.
We use an analytic method based on Descartes' rule of signs developed perviously for extremal RN black hole cases. Using this method, we first revisit the four-dimensional RN black hole case and obtain the same result as that in previous works.  We then study the five-dimensional non-extremal RN black hole case. It is found that all the five-dimensional non-extremal RN  black holes are superradiantly stable under charged massive scalar perturbation. The key point in the proofs is that there is no potential well for the effective potential felt by the superradiant modes of the scalar perturbation.

The two equations \eqref{4dsign} and \eqref{5dsign} play important roles in the proofs. In each case, it is unexpected that the possible signs of the complicated coefficients in the effective potential take  these interesting and simple forms. This does not seem like a coincidence. It will be interesting to extend the method here and try to prove that the non-extremal RN black holes are superradiantly stable under charged massive scalar perturbation in arbitrarily higher dimension.

\begin{acknowledgements}
This work is partially supported by Guangdong Major Project of Basic and Applied Basic Research (No. 2020B0301030008), Science and Technology Program of Guangzhou (No. 2019050001) and Natural Science Foundation of Guangdong Province (No. 2020A1515010388, No. 2020A1515010794).
\end{acknowledgements}

\end{document}